%% file: ltexpprt.tex
\documentclass[twoside,leqno,twocolumn]{article}

\usepackage[letterpaper]{geometry}
\usepackage{ifthen}
\usepackage{ltexpprt}
\usepackage{subfigure}
\newcommand{\eat}[1]{}
\usepackage{cite}
\usepackage{float}
\usepackage{amsmath,amssymb,amsfonts}
\usepackage{times}
\usepackage{algorithmic}
\usepackage{graphicx}
\usepackage{textcomp}
\usepackage{color}
\usepackage[table]{xcolor}
\usepackage{url}
\usepackage{ulem}

\def\BibTeX{{\rm B\kern-.05em{\sc i\kern-.025em b}\kern-.08em
    T\kern-.1667em\lower.7ex\hbox{E}\kern-.125emX}}

\newcommand{\showReminders}{yes}
\newcommand{\showComments}{yes}
\newcommand{\submit}{no}            

\newcommand{\note}[2]{
  \ifthenelse{\equal{\submit}{yes}}{}{%
    \ifthenelse{\equal{\showComments}{yes}}{\textcolor{#1}{#2}}{}
  }
}

\newcommand*{\dif}{\mathop{\mathrm{d}}\!}
\newcommand{\reminder}[1]{{{\bf [[#1]]}}}
\newcommand{\eatreminders}{\renewcommand{\reminder}[1]{}}
\ifthenelse{\equal{\submit}{yes}}{\eatreminders}{}
\ifthenelse{\equal{\showReminders}{yes}}{}{\eatreminders}

\usepackage{url}
\usepackage[pdftex]{hyperref}

\newtheorem{definition}{Definition}

\begin{document}

\title{\Large It's the Best Only When It Fits You Most: Finding Related Models for Serving \\Based on Dynamic Locality Sensitive Hashing  
}

\author{Lixi Zhou\thanks{lixi.zhou@asu.edu, Arizona State University}
\and Zijie Wang \thanks{zijiewang@asu.edu, Arizona State University}
\and Amitabh Das \thanks{adas59@asu.edu, Arizona State University}
\and Jia Zou\thanks{jia.zou@asu.edu, Arizona State University}}

\date{}

\maketitle


\fancyfoot[R]{\scriptsize{Copyright \textcopyright\ 20XX by SIAM\\
Unauthorized reproduction of this article is prohibited}}





\begin{abstract}
In recent, deep learning has become the most popular direction in machine learning and artificial intelligence. However, preparation of training data is often a bottleneck in the lifecycle of deploying a deep learning model for production or research.
Reusing models for inferencing a dataset can greatly save the human costs required for training data creation. Although there exist a number of model sharing platforms such as TensorFlow Hub, PyTorch Hub, DLHub, most of these systems require model uploaders to manually specify the details of each model and model downloaders to screen keyword search results for selecting a model. They are in lack of an automatic model searching tool. This paper proposes an end-to-end process of searching related models for serving based on the similarity of the target dataset and the training datasets of the available models. While there exist many similarity measurements, we study how to efficiently apply these metrics without pair-wise comparisons and evaluate the effectiveness of these metrics. We find that our proposed adaptivity measurement which is based on Jensen-Shannon divergence, is an effective measurement, and its computation can be significantly accelerated by using the technique of locality sensitive hashing.

\end{abstract}
\input{intro}

\input{background}

\input{solution}

\input{evaluation}
\input{relatedworks}
\section{Conclusions}
In this work, we systematically explore the problem of finding related models for serving based on JS-divergence and adaptivity, which are dynamically computed over the features shared by the source and target domains.
We propose and implement an end-to-end system, called ModelHub, to efficiently automate the process of finding related models for serving based on various techniques including LSH for JS-divergence, the compressed and indexed probability distribution structure, and MinHash. We demonstrate the accuracy and efficiency of the proposed system through extensive experiments on a number of workloads including activity recognition, entity matching, image recognition, and natural language processing. Our proposed work can greatly save the human costs and shorten the deep learning model deployment time for production and research.

\section{Acknowledgment}
We would like to thank Dunchuan Wu for his help with Fig.~\ref{fig:tfhub}.

\bibliographystyle{abbrv}
\bibliography{refs}  

\end{document}

%% file: intro.tex
\section{Introduction}

Pre-trained models can be reused directly or through transfer learning so that the substantial efforts required for collecting and annotating training data on the new problems can be greatly saved~\cite{yang2017deep}. This has motivated a number of model sharing platforms, such as TensorFlow Hub~\cite{tensorflowhub}, PyTorch Hub~\cite{pytorchhub}, and DLHub~\cite{dlhub}. However, the selection of proper models for serving is mostly through the intuition of domain experts and trial-error processes. Consider the following example: \textit{In TensorFlow Hub~\cite{tensorflowhub}, there are now $134$ text embedding models available for downloads. Supposing we want to select one of these models to work on integrated open tables regarding coronavirus disease 2019 (COVID-19), which model can achieve optimal accuracy in the target domain and should be chosen?}

\noindent
To address such problem, most of the model sharing platforms like DLHub~\cite{dlhub} provides a model search tool based on the indexing of model metadata that mostly covers: (1) publication
schemas such as author, creation date, description;
(2) model information such as algorithm, software version, network architecture;
(3) development provenance such as versions, contributors;
(4) training information such as training datasets and training parameters;
(5) performance information such as accuracy on benchmark datasets.

However, none of this information can directly tell \textit{which model should be selected for inference on a target dataset}. The prediction of the serving accuracy of a model in a target dataset is inherently a complex problem. It is related not only to the model architecture, but also to the similarity and relationships between the source and target datasets. The model is the best only if it fits your dataset~\cite{ding2018model}. However, we observe that existing model serving platforms are lack of the capability to automatically search for a model that is trained on the best fitting dataset. People need to perform a text-based search using keywords, scrutinize each search result, and manually check the similarity of each training dataset to their own dataset. Such human-centered model selection is inefficient, which delays the model deployment and incurs significant human costs. To alleviate this problem, it is critical to automate the model selection process for model serving. The technical challenges include:

\noindent
(1) Storing training datasets with the models incurs significant storage overheads and privacy concerns.

\begin{figure}[h]
\centering
   \includegraphics[width=3.3in]{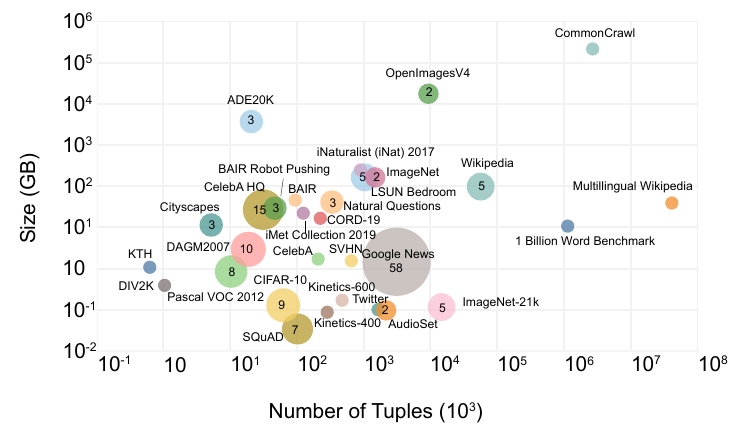}
\caption{\label{fig:tfhub}
TensorFlow Hub Models. The value and size of the circle denote the number of models trained on the dataset.}
\vspace{-5pt}
\end{figure}

\noindent
(2) The similarity measurements with all candidate datasets are time-consuming. The computational overhead not only scales to the number of available models in a domain, but also scales to the sizes of the training datasets to be compared. Taking TensorFlow Hub for example, text embedding domain has $134$ models, and image feature vector domain has $130$ models. In addition, the datasets used by models in TensorFlow Hub may have up to several hundreds of terabytes in size, as illustrated in Fig.~\ref{fig:tfhub}, which also significantly slows down similarity measurements. 


In this work, we address these two problems using locality sensitive hashing (LSH) for the probability distributions~\cite{chen2019locality, mao2017s2jsd}. While LSH is a popular technique to solve nearest neighbor problems, most of existing LSH families, are designed to measure the Hamming distance~\cite{datar2004locality} and Euclidean distance~\cite{indyk1998approximate} of two vectors with fixed dimensions, or the Jaccard similarity~\cite{broder1997resemblance} and cosine similarity of two high-dimensional binary vectors~\cite{charikar2002similarity}. Until very recently, the LSH for measuring similarity of probability distributions is studied~\cite{chen2019locality, mao2017s2jsd}. S2JSD-LSH~\cite{mao2017s2jsd} is first proposed to provide LSH functions for a new measurement $S2JSD_{new}^{approx}$, which is obtained by only keeping the linear terms in the Taylor expansion of the logarithm in the expression of $S2JSD$, the square root of two times the Jensen-Shannon (JS)  divergence. However, their LSH design doesn't provide any bound on the actual JS-divergence. Then Chen et al.~\cite{chen2019locality} propose a new LSH scheme for the generalized JS-divergence through the approximation of the squared Hellinger distance, which is in the form of the L2-LSH family~\cite{datar2004locality}, and called as JS-LSH in this paper. 

A significant problem is that JS-divergence is symmetric, however the serving accuracy may not be symmetric. For example, a dataset $\mathbf{A}$ may contain only a subset of labels of another dataset $\mathbf{B}$; then supposing we have a model $M_{\mathbf{A}}$ that is trained on $\mathbf{A}$, and a model $M_{\mathbf{B}}$ that is trained on $\mathbf{B}$, the accuracy of $M_{\mathbf{A}}$ on $\mathbf{B}$, and the accuracy of $M_{\mathbf{B}}$ on $\mathbf{A}$ could be very different. Obviously, simply applying the symmetric JS-divergence to the model selection problem may be problematic. While there exist asymmetric measurements of two probability distributions, such as Kullback-Leibler (KL) divergence, there doesn't exist efficient LSH families for these metrics. To address the problem, we propose to split a dataset into a number of partitions and compute the JS-LSH for each partition, so that each dataset is associated with multiple LSH signatures. Thus we propose a new metric, called adaptivity, based on the number of equivalent LSH signatures shared by two datasets and the size of the target dataset, which indicates the portion of samples in the target dataset that have been seen by the model.

Another problem is that before we can apply the adaptivity measurement for fast nearest neighbor search, we need first determine the overlapping feature space between the source domain and the target domain, because the features in the source domain may be very different from the target domain. It is straightforward to use MinHash~\cite{zhu2016lsh}, another LSH technique based on Jaccard similarity, to find all of the shared features between two datasets. However, because the overlapping space can only be determined at the query time, there is no way to pre-compute the JS-LSH signatures. To address the problem, we propose to use a compressed and indexed probability distribution to represent each model's training dataset by encoding occurrences of feature values falling into each hash bin, so that the storage overhead and runtime JS-LSH computation time can be significantly reduced without losing the effectiveness.

 \eat{First, a model is uploaded to \textit{ModelHub} with the MinHash signatures computed for each of features, and the compressed and indexed probability distributions pre-computed for the training dataset. The MinHash signatures are used to determine the overlapping in feature space between source and target based on Jaccard similarity. In this process, each feature that contains numerical values are first transformed into histograms. Then a user, who wants to transfer a pre-trained model on his target dataset, will issue a query to identify the most related model. The query must contain the Minwise LSH signatures for each of the features as well as the JS-divergence's $L^2LSH$ signatures extracted from the user's target dataset. Then the \textit{TransferHub} system will first match the query to a cluster of models based on the Minwise hashes, and then match the query to a model in the selected cluster using the $L^2LSH$ signatures for JS-divergence.}
Our main contributions are summarized as following:

\noindent
(1) As far as to our knowledge, we are the first to systematically explore the problem of finding related models for serving a target domain by comparing various metrics. 

\noindent
(2) We propose an adaptivity metric and an end-to-end system, called as \textit{ModelHub}, to automate the process of finding related models for serving based on dynamic computation over the features shared by the source and target domains using LSH techniques.

\noindent
(3) We conduct extensive experiments to compare the effectiveness of the proposed system to alternative solutions.

%% file: background.tex
\section{Background}
\subsection{Jensen-Shannon (JS) Divergence}

JS-divergence~\cite{lin1991divergence} is a measurement of the similarity of two probability distributions, which is a symmetric metric derived from the asymmetric Kullback-Leibler (KL) divergence~\cite{kullback1951information}.

Let $P$ and $Q$ be two probability distributions associated with a common sample space $\Omega$, and let $M = (P+Q)/2$. The JS-divergence is defined by:

\begin{equation}
    D_{JS}(P \parallel Q) = \frac{1}{2}D_{KL}(P \parallel M) + \frac{1}{2}D_{KL}(Q \parallel M)
\end{equation}

Here, $D_{KL}$ denotes the Kullback-Leibler divergence, which is defined as following for discrete distributions:

\begin{equation}
    D_{KL}(P \parallel Q) = \sum_{\textbf{x} \in \Omega}{P(\textbf{x} ) \log (\frac{P(\textbf{x} )}{Q(\textbf{x})})},
\end{equation}

\noindent
and as following for continuous distributions:

\begin{equation}
    D_{KL}(P \parallel Q) = \int_{\Omega}{ P(\textbf{x} )\log (\frac{P(\textbf{x} )}{Q(\textbf{x} )})\dif{\textbf{x} }}
\end{equation}

In this work, we mainly consider to leverage JS-divergence to select models for serving. First, JS-divergence is a widely used similarity metric for probability distributions~\cite{lee2000measures}. Second, it is easier to find LSH schemes for JS-divergence, compared to other similarity measurements of probability distributions~\cite{mao2017s2jsd}.

While other metrics such as Maximum Mean Discrepancy (MMD) could measure the domain adaptivity more accurately, it requires an optimization process and significantly higher computational costs, as illustrated in Tab.~\ref{tab:latency}.

\begin{table}
\centering
\scriptsize
\caption{\label{tab:latency} Average latency comparison on Activity Recognition (Sec.~\ref{workloads}) datasets (Unit: seconds).}
\begin{tabular}{|r|r|r|r|} \hline
Jaccard similarity&KL-divergence&JS-divergence (w/o LSH)&MMD\\\hline \hline
$379$&$29$&$87$&$6451$\\ \hline
\end{tabular}
\vspace{-5pt}
\end{table}

\subsection{Locality Sensitive Hashing (LSH)}
LSH was developed for general approximate nearest neighbor search problem~\cite{indyk1998approximate}. It requires a family of LSH functions, each of which is a hash function whose collision probability increases with the similarity of the inputs. The $(r_1, r_2, p_1, p_2)$-sensitive LSH family is formally defined in Definition.~\ref{def1}.

\begin{definition}
\label{def1}
Let $\mathcal{F}=\{h: M \rightarrow U\}$ be a family of hash functions for distance measurement $D$. $\mathcal{F}$ is $(r_1, r_2, p_1, p_2)$-sensitive  $(r_1 < r_2$ and $p_1 > p_2)$, if $\forall p, q \in M$, it satisfies that: (1) if $D(p, q) \leq r_1$, we have $Pr[h(p) = h(q)] \geq p_1$; (2) if  $D(p, q) \geq r_2$, we have $Pr[h(p) = h(q)] \leq p_2$.
\end{definition}

LSH is first proposed by Indyk and et al~\cite{indyk1998approximate} for measuring the Hamming distance in a $d$-dimensional Euclidean space, which requires the data to be vectors with fixed dimensions. MinHash~\cite{broder1997resemblance} is a widely used family of hash functions for Jaccard similarity. SimHash is an LSH scheme for Cosine distance~\cite{charikar2002similarity}. Both of MinHash and SimHash are only applicable to high-dimensional binary vectors or sets of values (without fixed dimensions), and MinHash is usually considered to be more computationally efficient than SimHash~\cite{zhu2016lsh}. However none of these popular LSH schemes are applicable to our problem where each dataset is a variable number of high-dimensional vectors.

As mentioned, the LSH schemes for probability distributions are recently studied~\cite{chen2019locality, mao2017s2jsd}. S2JSD-LSH~\cite{mao2017s2jsd}  provides LSH functions for a measurement that approximates the square root of two times the JS-divergence. Unfortunately, their LSH design doesn't provide any bound on the actual JS-divergence. Then Chen et al.~\cite{chen2019locality} propose a new LSH scheme for the generalized JS-divergence through the approximation of the squared Hellinger distance, which is proved to be bounded with the actual JS-divergence and thus used in this work as JS-LSH, as defined in Eq.~\ref{eq:eq1}:

\begin{equation}
h_{\textbf{a},b}=\lceil{\frac{\textbf{a} \cdot \sqrt{P}+b}{r}}\rceil,
\label{eq:eq1}
\end{equation}

\noindent
where $P$ is a probability distribution in the sample space $\Omega$, $\textbf{a} \sim \mathcal{N}(0, I)$ is a $|\Omega|$-dimensional standard normal random vector, $\cdot$ represents the inner product operation, $b$ is a random variable uniformly distributed on $[0, r]$, and $r$ is a positive real number. This approximation is proved to be lower bounded by a factor $0.69$ for the JS-divergence~\cite{chen2019locality}. 

 \eat{
\subsection{Transfer Learning} 

Transfer learning is the process that recognizes and applies the knowledge learnt in a previous task (i.e., the source task) to new tasks (i.e., the target task)~\cite{pan2009survey}. The source task is denoted as $\mathcal{T}_S = \{\mathcal{Y}_S, f_S(\cdot)\}$, where $\mathcal{Y}_S$ defines the label space, and $f_S(\cdot)\}$ defines the label prediction function, which is equivalent to $P(\mathcal{Y}_S|\mathcal{X}_S)$. The target task is correspondingly represented as $\mathcal{T}_T= \{\mathcal{Y}_T, f_T(\cdot)=P(\mathcal{Y}_T|\mathcal{X}_T)\}$. In addition, the source domain, denoted as $\mathcal{D}_S = \{\mathcal{X}_S, P(\textbf{x})\}$, is composed of the feature space $\mathcal{X}_S$ and the probability distribution $P(\textbf{x}), \textbf{x}=(x_1, x_2, ..., x_n) \in \mathcal{X}_S$, on which the source task is trained. Similarly, a target domain is defined as $\mathcal{D}_T = \{\mathcal{X}_T, P(\textbf{x})\}, \textbf{x}=(x_1, x_2, ..., x_n) \in \mathcal{X}_T$. The transfer learning process that we are focused on in this work can be formalized in Definition~\ref{def2}.

\begin{definition}
\label{def2}
Given a source domain $\mathcal{D}_S$ and source task $\mathcal{T}_S$, a target domain $\mathcal{D}_T$ and target task $\mathcal{T}_T$, transfer learning targets at learning the parameters to optimize the objective function $f_T(\cdot)$ on $\mathcal{D}_T$, using knowledge learnt in $\mathcal{T}_S$ on $\mathcal{D}_S$, where $\mathcal{D}_S$ and $\mathcal{D}_T$ are different but related, and/or $\mathcal{T}_S$ and $\mathcal{T}_T$ are different but related.
\end{definition}

There are many techniques for transfer learning~\cite{pan2009survey, weiss2016survey, zhuang2020comprehensive}, which are usually categorized as: (1) Instance-based transfer, which is to re-weight some labeled data in the source domain and use these data to train the model in the target domain~\cite{}; (2) Feature-based transfer, which is to transform the source and target feature spaces so that the differences between the two domains can be minimized~\cite{}; (3) Model/parameter-based transfer, which is to identify the shared parameters or priors between the source and target models, freeze these parameters, and only fine-tune non-shared parameters using a small amount of labeled data from the target domain~\cite{}; (4) relational-knowledge-based transfer, which is to build mapping of relational knowledge between the source domain and the target domain, then the source model will be revised using the relational mapping for inferrence on the target domain~\cite{}. 

In this work, we mainly focus on the two most popular types of techniques: feature-based transfer learning and model-based transfer learning.

}

%% file: solution.tex
\section{Methods and Technical Solutions }
\subsection{A New Metric Derived from JS-divergence}
As mentioned, JS-divergence is a symmetric measurement, while the model adaptivity between two datasets is usually asymmetric. Considering a dataset $\mathbf{A}$ that contains a subset of dataset $\mathbf{B}$, then the model trained on $\mathbf{B}$ may have good accuracy on $\mathbf{A}$, because the model has seen all of the samples in $\mathbf{A}$ during the training process. However, the model trained on $\mathbf{A}$ may have relatively worse accuracy on $\mathbf{B}$ if the probability distributions of $\mathbf{B}$ and $\mathbf{A}$ are not identical. While asymmetric measurements such as KL-divergence can address the issue, there is no existing work that has discussed the LSH for measuring KL-divergence. In this work, we propose a new measurement, called adaptivity, to solve the problem by randomly partitioning $\mathbf{A}$ and $\mathbf{B}$ into partitions and computing a ratio of the number of similar partitions to the total number of partitions in the target dataset, as defined in Def.~\ref{new-metric}.
\begin{definition}
\label{new-metric}
Supposing we have two datasets  $\mathcal{D}_s = \{p_0, ..., p_{ns}\}$ and $\mathcal{D}_t = \{q_0, ..., q_{nt}\}$, where $p_i (0 \leq i < ns)$ and $q_j (0 \leq j < nt)$ are partitions that have equivalent sizes (except for the last residue partition), given a threshold $t$, we can join $\mathcal{D}_s$ and $\mathcal{D}_t$ to identify all pairs in form of $(p_i, q_j)$, so that $D_{JS}(p_i, q_j) \leq t$.  We denote the total number of pairs that satisfy above condition as $num\_matches$. Then we can derive a new metric to measure the adaptivity of the models trained in the source domain to the target domain, denoted as: $adaptivity(\mathcal{D}_s, \mathcal{D}_t) = \frac{num\_matches}{nt}$.
\end{definition}

\subsection{Problem Formulation}
We divide the problem of selecting a model for reuse in two sub-problems. The first sub-problem is to obtain the overlapping features of the training data (i.e. source domain) and the query data (i.e. target domain). The second sub-problem is to measure the adaptivity over the overlapping feature space. The two sub-problems are formulated as following.

\begin{definition}
\label{def3}
Suppose we have a database of $n$ pre-trained models (i.e., source tasks in a source domain), represented as $\mathcal{M}=\{M_1, ..., M_n\}$. Each model $M_i \in \mathcal{M}$ has $k_i$ features, denoted as $F_i=\{f_1, ..., f_{k_i}\}$. Given a query $q$ that has $k_q$ features, denoted as $F_q=\{f_1, ..., f_{k_q}\}$, and two thresholds $t_1, t_2 \in [0,1]$, the database should return a set of relevant models, denoted as $\mathcal{M}_q \subset \mathcal{M}$, so that $\forall M_i \in \mathcal{M}_q, overlap(F_i, F_q) = \frac{F_i \cap F_q}{F_q} > t_1$. Two features $f_i$ and $f_j$ are equivalent if and only if $jaccard(f_i, f_j) > t_2$.
\end{definition}

\begin{definition}
\label{def2}
Supposing each model $M_i \in \mathcal{M}$ has a source domain represented as $\mathcal{D}_i$. Given a query that has a target domain $\mathcal{D}_q$, as well as a threshold $t_{adaptivity}$ or $t_{JS}$, depending on which metric is used, the database should return a set of relevant models, denoted as $\mathcal{M}_q \subset \mathcal{M}$, so that $\forall M_i \in \mathcal{M}_q, adaptivity(\mathcal{D}_i, \mathcal{D}_q) \geq t_{adaptivity}$ or $D_{JS}(\mathcal{D}_i, \mathcal{D}_q) \leq t_{JS}$. Here, $\mathcal{D}_i$ and $\mathcal{D}_q$ are defined over the overlapped feature space as defined in Def.~\ref{def3}.
\end{definition}

\subsection{Solution Architecture}
We propose ModelHub as an end-to-end solution to address the above two problems. ModelHub is a model sharing platform, which requires that for each model, following information must be provided:

\noindent
(1) A MinHash signature for each feature in the training data. MinHash is a widely used technique to efficiently search nearest neighbors based on the Jaccard similarity. For a feature that contains continuous numerical values, we first quantize the values into a number of bins based on the range and distribution of data, to improve the effectiveness of Jaccard similarity estimation for such features~\cite{wu2020review}.

\begin{figure}[h]
\centering
   \includegraphics[width=3in]{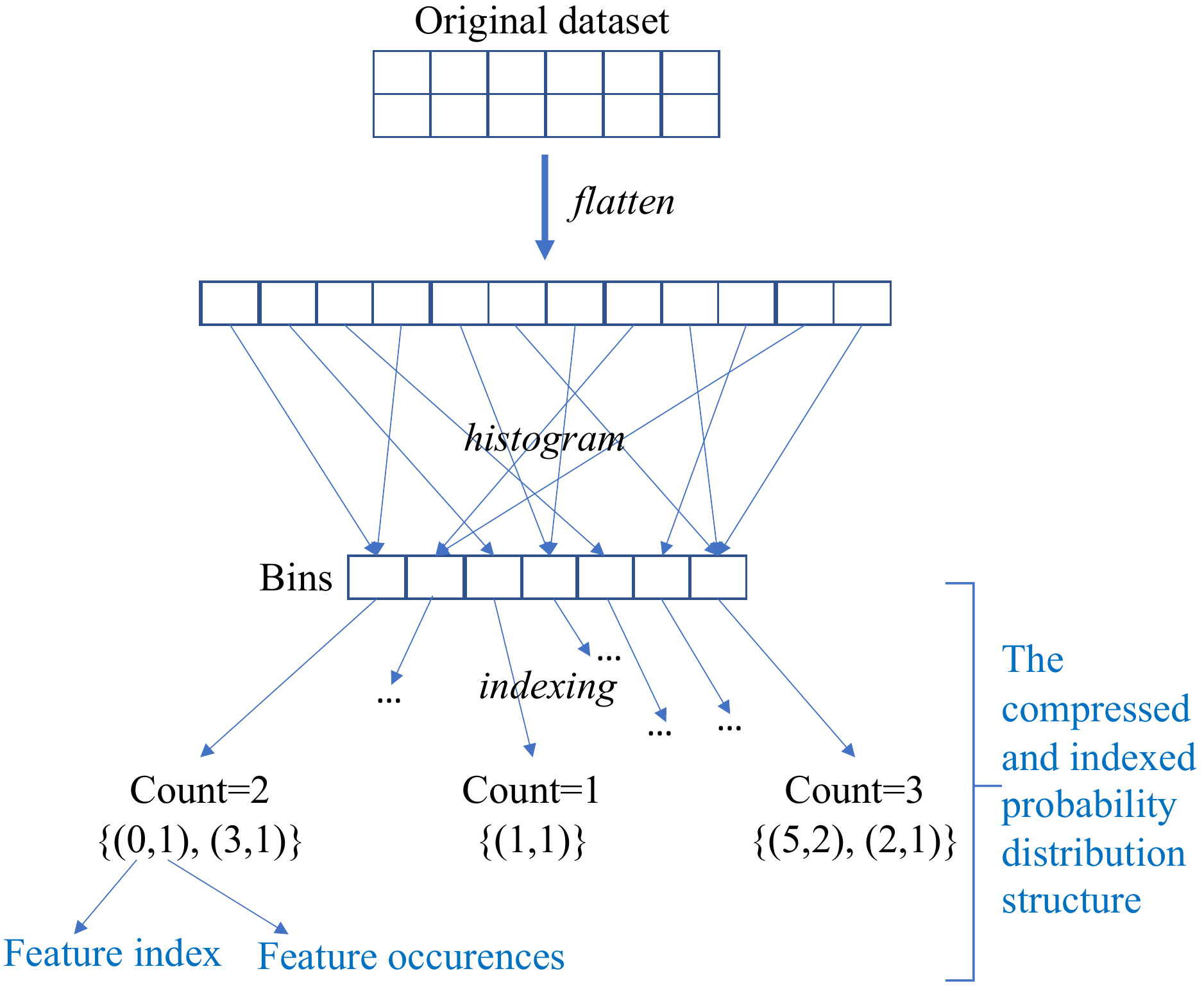}
\caption{\label{fig:structure}
Succinct representation of a (partition of) dataset}
\end{figure}

\noindent
(2) A compressed data sketch for each partition of the training dataset, which can be used to easily derive the probability distribution over any subset of features in this partition. This structure is used to derive the JS-LSH signatures for efficiently measuring JS-divergence in the dynamic overlapping feature space shared by the training dataset of this model (i.e. the source domain) and the querying dataset (i.e. the target domain). Most training datasets are multi-dimensional, and to compute the multi-variate probability distribution of a multi-dimensional dataset is difficult~\cite{statisticsch3}. To solve the problem, we first flatten a multi-dimensional dataset into a single-dimensional dataset~\cite{chen2019locality} and then create a set of histogram bins over the flattened dataset. Finally, we create an index, each entry contains a feature ID followed by the feature's occurrences in each bin, as illustrated in Fig.~\ref{fig:structure}. All partitions across all datasets have equivalent size $m$.

Based on these information, the system can efficiently compute the adaptivity based on Def.~\ref{new-metric}, without uploading the training dataset. The process is depicted as following.

\noindent
Step 1. Given a query dataset, we need first determine all models that have significant overlaps in feature space with the query dataset using the MinHash  signatures. (Sec.~\ref{sec:overlap})

\noindent
Step 2. For each of the models returned from Step 1, we use adaptivity to select the best model. \eat{For JS-divergence, we simply compute the $L^2LSH$ hash signature (as illustrated in  Eq.~\ref{eq:eq1}) from the compressed structure of each candidate model, and compare these with the $L^2LSH$ computed from the query dataset. If adaptivity is used,} We compute the JS-LSH signatures over the shared features for each partition, store the signatures of the larger dataset that has more partitions to hash tables, and then use the signatures of the smaller dataset to look up in the hash tables to determine the number of similar partitions. Then we estimate the adaptivity of the source domain and the target domain, as defined in Def.~\ref{new-metric} and Def.~\ref{def2}.(Sec.~\ref{sec:js})

\subsection{Overlap Search}
\label{sec:overlap}
The search of the overlapping features of a source dataset and the target dataset are based on MinHash~\cite{broder1997resemblance}, a widely used locality sensitive hashing technique for Jaccard similarity. The search process contains an offline and an online stage. We first describe the offline stage. First, for each model, we expand each feature into a vector of values based on its succinct representation of the training data,  by replicating the value associated with each bin by the number of its occurrences. Then for each feature vector, we will create $L_{minhash}$ MinHash signatures corresponding to $L_{minhash}$ bands. Each band is associated with $K_{minhash}$ MinHash functions, and each signature is a concatenation of $K_{minhash}$ MinHash values produced by the band's functions. A hash table is created for each band, so that for each model, each of the $L_{minhash}$ signatures is stored to the corresponding hash table.

Then at the online stage, a model search request regarding a target dataset is issued. The target dataset's MinHash signatures will be computed for each feature on each band using the aforementioned approach. Then for each feature, the signature of each band will be used to query the corresponding hash table to find out all models that have a matching signature. Finally,  models having a significant portion of features overlapping with the target dataset, as defined in Def.~\ref{def3}, will be identified and returned.

\subsection{Model Search}
\label{sec:js}
In this section, we mainly describe how the adaptivity metric is computed over the identified overlapping feature space at runtime. First, for each partition in the training dataset of a model $M_i$,  we expand the succinct representation into a one-dimensional dataset that concatenates all expanded feature vectors. Then for each partition's dataset, we create $L_{jslsh}$ bands, and each band is associated with $K_{jslsh}$ hash functions based on Eq.~\ref{eq:eq1}, we create a JS-LSH signature of length $K_{jslsh}$ by concatenating values computed by these hash functions. Each band is constructed with a hash table and each signature is stored into the corresponding band's hash table. 

\eat{
First, for each partition in the model $M_i$, we create an JS-LSH signature by concatenating values computed using a family of $L_{l2hash}$ hash functions based on Eq.~\ref{eq:eq1}. If the training dataset of $M_i$ contains $n_i$ samples, it will have $n_i/m$ signatures, where $m$ is the size of each partition, and each signature has the size of $L_{l2hash}$. Following that, we split each signature into $K_{l2hash}$ bands and store the parts of all signatures fall into the same band into the same hash map out of $K_{l2hash}$ hash maps.
}

Then, we do the same for the target dataset to obtain $L_{jslsh}$ signatures for each band in each partition, which are used to query the corresponding hash table to find all models, of which a partition of the training data has a matching signature in the same band. Based on the matches, we can compute the adaptivity as defined in Def.~\ref{new-metric}. Finally, we will select the models whose adaptivity with the target dataset passes a threshold $t_{adaptivity}$ as described in Def.~\ref{def2}. The computation of JS-divergence is a special case of the above process where the number of partitions is set to $1$.

\eat{ 
we create an $L^2LSH$ signature using a similar approach. The $L^2LSH$ signature will also be split into $K_{l2hash}$ bands. Each band is used to query the corresponding hash map and find all source datasets that match that band.
}

%% file: evaluation.tex
\section{Empirical Evaluation}
\subsection{Workloads and Datasets}
\label{workloads}
We evaluate our proposed methodology using four workloads.

\noindent
\textbf{1. Activity Recognition.} Human activity recognition (HAR), is to predict the activities (e.g. walking, sitting, running, lying) based on data collected from multiple sensors attached to human body. HAR is a hot research topic in the pervasive computing area, and has been widely applied to indoor localization, sleep state detection, smart home sensing, and virtual reality~\cite{avci2010activity}. In this work, we use three public activity recognition datasets, including OPPORTUNITY~\cite{chavarriaga2013opportunity}, PAMAP2~\cite{reiss2012introducing}, and UCI DSADS~\cite{barshan2014recognizing}. Each dataset is extracted with $81$ features~\cite{wang2018stratified}. The OPPORTUNITY dataset is collected from four human subjects executing various activities with sensors attached to more than five body parts. The PAMAP2 dataset is collected from nine subjects performing $18$ activities with sensors attached to three body parts. The UCI DSADS dataset is collected from eight subjects wearing sensors on five body parts. We create $162$ tables by sampling these datasets, so that each table represents samples collected from a body part for specific subject. Then we apply our approach to a series of scenarios where models are trained on each of these tables, and given a target dataset collected from a subject on a specific body part, we want to select a model to achieve the best accuracy on the target dataset. 

\noindent
\textbf{2. Entity Matching (EM).} Entity matching  is to tell whether two tuples are referring to the same entity~\cite{li2020deep, zhao2019auto, ebraheem2017deeper, cappuzzo2020creating, fernandez2018seeping}. \eat{Many recent EM tools are based on deep learning~\cite{mudgal2018deep, zhao2019auto, ebraheem2017deeper, zhang2020multi}.} We mainly apply DeepMatcher~\cite{mudgal2018deep}, which is an EM tool based on deep learning, to four datasets~\footnote{https://github.com/anhaidgroup/deepmatcher/blob/master/Datasets.md}: Walmart-Amazon, Abt-Buy, DBLP-Scholar, DBLP-ACM; and $11$ smaller datasets from the Magellan Data Repository~\cite{magellandata}~\footnote{https://sites.google.com/site/anhaidgroup/useful-stuff/data}.   Each task contains training and testing samples collected from two different datasets. For example, IMDB-TMD is to match movie tuples collected from IMDB and TMD respectively. \eat{and IMDB-RottenTomatoes is to match movie tuples from IMDB and Rotten Tomatoes respectively.} Each EM task may have different features, and some tasks share a significant portion of common features. We focus on the model selection scenarios where a dataset is used for query, and a set of models are trained for the EM tasks on the rest of the datasets as candidates, and then we try to select a candidate model to serve the query dataset.

\noindent
\textbf{3. Image Recognition.} To create five image datasets with distinct probability distributions, we randomly sample images from the Cifar-10 dataset without replacement. Cifar-10 is a collection of images with labels of ten classes, each containing $5,000$ images. Then we make four of the subsets skewed by adding $5000$ images of the fourth class to the second and the fourth subset; $5000$ images of the second class to the third subset; $5000$ images of the eighth class to the third subset.  These five datasets are called as \texttt{Balanced}, \texttt{Skewed-1}, \texttt{Skewed-2}, \texttt{Skewed-3}, \texttt{Skewed-4} correspondingly. To evaluate the effectiveness of our proposed metrics for the image recognition scenario, we train ResNet56v1 model on each of the subsets respectively, and then perform five experiments of searching for the best model to be reused on each of the five datasets.

\noindent
\textbf{4. Natural Language Processing (NLP).} We are mainly interested in two types of NLP tasks. The first task is to identify whether sentence pairs have equivalent semantic meanings. For this task, we train models on three different datasets: Microsoft Research Paraphrase Corpus (MRPC) that includes $3,549$ samples; Quora Question Pairs (QQP) that consists of $363,192$ samples; and Paraphrase Adversaries from Word Scrambling (PAWS) which has $49,401$ samples. 
The second task is about natural language inference (NLI), which is to identify the textual entailment relationship~\cite{maccartney2009natural} between sentence pairs. Similar to the first task, we train models on three different datasets: Recognizing Textual Entailment (RTE) that has $2,490$ samples; Question NLI, which contains $510,711$ samples; and the SCITAIL dataset including $5,302$ samples, which is an entailment dataset created from multiple-choice science exams and web sentences.
We train models for these tasks using the same pre-trained BERT base model~\footnote{https://tfhub.dev/google/bert\_uncased\_L-12\_H-768\_A-12/1}, denoted as uncased\_L-12\_H-768\_A-12, which contains $12$ layers and $110$ millions of parameters. Then for each task, we use one of the three datasets as the target dataset, and try to choose the best model to serve on the target dataset.

\subsection{Evaluation Methodology} 
In these experiments, we mainly compare our proposed adaptivity metric (as defined in Def.~\ref{new-metric}) to other alternatives, including (1) JS-divergence (a special case of adaptivity when each dataset has only one partition), which is computed using the JS-LSH for JS-divergence; (2) L2-distance, which is to get the center of the overlapping features for the source and target datasets, and then compute the L2-LSH for Euclidean distance~\cite{indyk1998approximate}; (3) source accuracy, which is the testing accuracy of the candidate model on the source dataset.

To compare the effectiveness of these metrics, we compute the Pearson correlation coefficient values to show the impact of various metrics, including the adaptivity, the JS-divergence, the L2-distance, and the source accuracy, to the target accuracy (i.e. the prediction accuracy of the source model on the target dataset). In most of the cases, JS-divergence and L2-distance have the negative relationships with the target accuracy, so their Pearson correlation coefficient values are often negative.
In addition, we also evaluate and compare the error rate for each metric, while the top1-error and top2-error are defined as the number of wrong predictions in identifying the best model or one of the top two models to the total predictions.

\subsection{Results}

\subsubsection{Activity Recognition}
\begin{figure}[h]
\centering
   \includegraphics[width=3.3in]{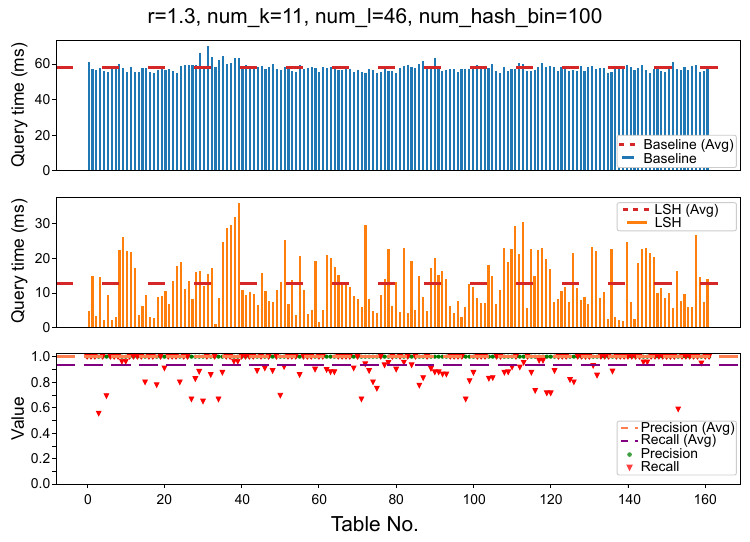}
\caption{\label{fig:ar-comparison}
Comparison of LSH for JS-divergence (denoted as LSH) to a pair-wise computation of JS-divergence (denoted as baseline) for $162$ activity recognition tables.}
\vspace{-5pt}
\end{figure}

We first compare the accuracy and latency of our proposed approach based on LSH for JS-divergence to a baseline approach that computes the JS-divergences between the target dataset and each of the source datasets directly without using LSH techniques. The results are illustrated in Fig.~\ref{fig:ar-comparison}, in which each bar represents a test case that uses one table as the query (i.e. target dataset), and the rest of the tables as the sources. In this experiment, for each test, we assume the source datasets that have the JS-divergence with the query dataset smaller than $0.1$ compose the ground truth. We observe that using LSH can significantly accelerate the overall latency required for JS-divergence comparison, and achieve $5\times$ speedup on average, while the precision of our proposed approach is $100\%$ and recall is above $93\%$, which are both acceptable results. 

\begin{figure}[H]
\centering
{\subfigure[r]{%
   \label{fig:r}
   \includegraphics[width=3.3in]{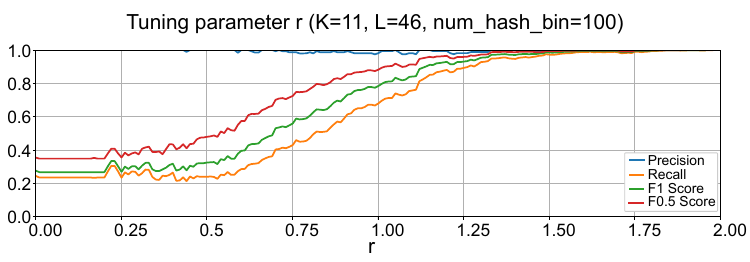}  
}%
\vspace{1pt}
\subfigure[number of concatenated hash functions]{%
  \label{fig:K}
  \includegraphics[width=3.3in]{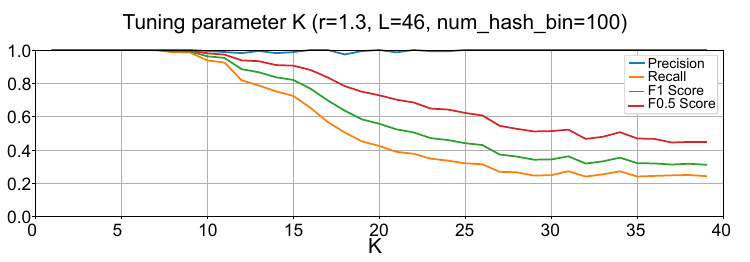}
}%
\vspace{1pt}
\subfigure[number of hash tables]{%
  \label{fig:L}
  \includegraphics[width=3.3in]{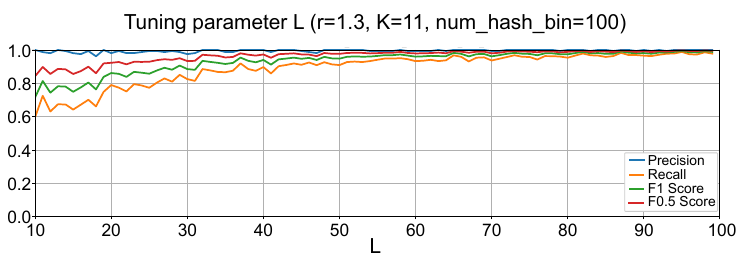}
}%
\vspace{1pt}
\subfigure[number of hash bins]{%
  \label{fig:L}
  \includegraphics[width=3.3in]{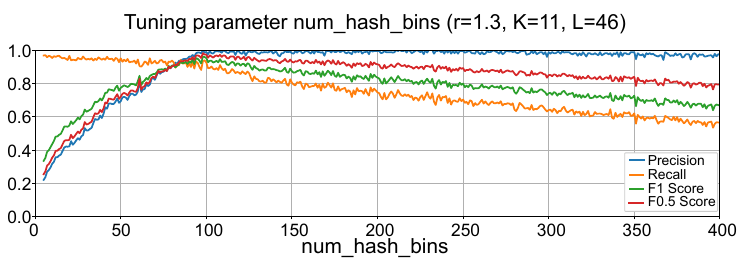}
}
}
\caption{\label{fig:ar-tuning}
Hyperparameter tuning for LSH of JS-divergence.
}

\end{figure}

We further evaluate how hyperparameters, such as $r$ (as in Eq.~\ref{eq:eq1}), the number of concatenated hash functions (i.e., $K_{l2hash}$) of each band, the number of bands (i.e., $L_{l2hash}$), and the number of hash bins, will affect the accuracy of the JS-divergence computed using LSH compared to the baseline, which are illustrated in Fig.~\ref{fig:ar-tuning}.

The comparison of Pearson correlation coefficient values of various metrics, including the adaptivity, the JS-divergence, the L2-distance, and the source accuracy, to the target accuracy are illustrated in Tab.~\ref{tab:comparison-to-model-adaptivity}. It shows that our proposed adaptivity metric can achieve better performance. Also among the four metrics, the source accuracy exhibits the least relevance with the target accuracy, which demonstrates our assumption that "It’s the best only when it fits you most", and indicates that the similarity between the source and the target plays a more important role in determining model adaptivity.

We also compute the top-1 and top-2 error rates, as illustrated in Tab.~\ref{tab:comparison-error-rate}.
It shows that all of the metrics including adaptivity, JS-divergence, and L2-distance can serve as a metric for finding related models with zero top-2 errors in our test cases. In addition, adaptivity and L2-distance have better top-1 error rate than JS-divergence.


\begin{table}
\centering
\scriptsize
\caption{\label{tab:comparison-to-model-adaptivity} Comparison of Pearson correlation coefficient for activity recognition (the best Pearson correlation coefficient values are highlighted in bold).}
\begin{tabular}{|l|r|r|r|r|} \hline
target&adaptivity&JS-divergence&l2-distance&source-accuracy\\\hline \hline
dsads\_la&$\textbf{0.61}$&$-0.17$&$-0.54$&$-0.44$\\ \hline
dsads\_ll&$\textbf{0.72}$&$-0.40$&$-0.68$&$-0.40$\\ \hline
dsads\_ra&$0.51$&$-0.13$&$-\textbf{0.54}$&$-0.37$\\ \hline
dsads\_rl&$0.58$&$-0.49$&$-\textbf{0.75}$&$-0.58$\\ \hline
dsads\_t&$0.19$&$-0.11$&$-0.42$&$-\textbf{0.44}$\\ \hline
oppo\_b&$\textbf{0.89}$&$-0.66$&$-0.77$&$0.45$\\ \hline
oppo\_lla&$\textbf{0.79}$&$-0.54$&$-0.54$&$0.34$\\ \hline
oppo\_lua&$\textbf{0.91}$&$-0.72$&$-0.76$&$0.48$\\ \hline
oppo\_rla&$\textbf{0.76}$&$-0.54$&$-0.52$&$0.28$\\ \hline
oppo\_rua&$\textbf{0.84}$&$-0.67$&$-0.70$&$0.44$\\ \hline
pamap\_a&$\textbf{0.69}$&$-0.68$&$-0.54$&$0.24$\\ \hline
pamap\_c&$\textbf{0.77}$&$-0.69$&$-0.70$&$-0.06$\\ \hline
pamap\_w&$\textbf{0.83}$&$-0.74$&$-0.63$&$0.14$\\ \hline
\end{tabular}
\vspace{-5pt}
\end{table}

\begin{table}
\centering
\scriptsize
\caption{\label{tab:comparison-error-rate} Comparison of error rate for activity recognition}
\begin{tabular}{|l|r|r|r|r|} \hline
          &adaptivity&JS-divergence&l2-distance&source-accuracy\\\hline \hline
top-1 error&$\textbf{23\%}$&$31\%$&$\textbf{23\%}$&$92\%$\\ \hline
top-2 error&$\textbf{0\%}$&$\textbf{0\%}$&$\textbf{0\%}$&$92\%$\\ \hline
\end{tabular}
\vspace{-5pt}
\end{table}

\subsubsection{Entity Matching}
In this experiment, different from activity recognition, most attributes contain text-based values. We represent the training dataset of each entity matching task as a bag of words over a shared dictionary for computing the JS-divergence and adaptivity. We compare the Pearson correlation coefficient values of the JS-divergence metric, the adaptivity metric, and source accuracy, to the target accuracy for each EM task, as illustrated in Tab.~\ref{tab:comparison-to-model-adaptivity-em}. We also compare the overall accuracy in terms of top1-error and top2-error for all fifteen tasks, as illustrated in Tab.~\ref{tab:comparison-error-rate-em}. The results show that the adaptivity metric outperforms other metrics in selecting the model to serve with the best accuracy.

\begin{table}
\centering
\scriptsize
\caption{\label{tab:comparison-to-model-adaptivity-em} Comparison of Pearson correlation coefficient for EM (the best Pearson correlation coefficient values are highlighted in bold)}
\begin{tabular}{|l|r|r|r|} \hline
target&adaptivity&JS-divergence&source-accuracy\\\hline \hline
Abt\_Buy&$\textbf{0.99}$&$-\textbf{0.99}$&$\textbf{0.99}$\\ \hline
Dplp\_Acm&$0.92$&$-\textbf{0.94}$&$0.70$\\ \hline
Dblp\_Scholar&$0.91$&$-\textbf{0.94}$&$0.83$\\ \hline
Walmart\_Amazon&$\textbf{0.99}$&$-\textbf{0.99}$&$0.73$ \\ \hline
MyAnimeList\_AnimePlanet&$\textbf{0.61}$&$-0.53$&$0.51$\\ \hline
Bikedekho\_Bikewale&$\textbf{0.51}$&$-0.16$&$0.19$\\ \hline
Amazon\_Barnes&$\textbf{0.24}$&$-0.17$&$0.22$\\ \hline
GoodReads\_Barnes&$\textbf{0.61}$&$-0.53$&$0.20$\\ \hline
Barnes\_Half&$0.29$&$-\textbf{0.44}$&$0.31$\\ \hline
RottenTomatoes\_IMDB&$0.11$&$-\textbf{0.34}$&$0.16$\\ \hline
IMDB\_TMD&$0.49$&$-\textbf{0.67}$&$0.44$\\ \hline
IMDB\_RottenTomatoes&$0.51$&$-\textbf{0.75}$&$0.22$\\ \hline
Amazon\_RottenTomatoes&$0.60$&$-\textbf{0.64}$&$0.38$\\ \hline
RogerElbert\_IMDB&$\textbf{0.58}$&$-0.13$&$0.07$\\ \hline
YellowPages\_Yelp&$\textbf{0.82}$&$-0.67$&$0.44$\\ \hline
\end{tabular}
\vspace{-5pt}
\end{table}

\begin{table}
\centering
\scriptsize
\caption{\label{tab:comparison-error-rate-em} Comparison of error rate for activity recognition}
\begin{tabular}{|l|r|r|r|} \hline
&adaptivity&JS-divergence&source-accuracy\\\hline \hline
top-1 error&$\textbf{13\%}$&$47\%$&$33\%$\\ \hline
top-2 error&$\textbf{0\%}$&$27\%$&$27\%$\\ \hline
\end{tabular}
\vspace{-5pt}
\end{table}

\subsubsection{Image Recognition}
In this scenario, for each Cifar-10 image, we normalize all pixel values by dividing each value by $255$, and then flatten the image from a tensor of the shape ($32$, $32$, $3$) into a one-dimensional vector of the shape ($1$, $3072$). Then, we compute the JS-divergence and L2-distance on the flattened dataset. \eat{As mentioned in Sec.~\ref{workloads}, we randomly create five datasets, each has a different probability distribution. Only one dataset, which contains $5000$ images, has uniform distribution of image classes. Each of the four other datasets consists of $11,000$ images, and is skewed in one of the classes. As mentioned in Sec.~\ref{workloads}, we conduct five experiments, and in each experiment, one of the five datasets is considered as the target dataset and we try to use different metrics to select a model that is trained on one of the rest of the datasets to serve on the target dataset.}

We evaluate and compare the effectiveness of adaptivity, JS-divergence, L2-distance, and source accuracy for searching the best model to serve on a target dataset. The Pearson correlation coefficient of each metric to the target accuracy in each experiment is illustrated in Tab.~\ref{tab:comparison-to-model-adaptivity-image}. The top-1 and top-2 error rate for searching the best model for each experiment is illustrated in Tab.~\ref{tab:comparison-error-rate-image}.

\begin{table}
\centering
\scriptsize
\caption{\label{tab:comparison-to-model-adaptivity-image} Comparison of Pearson correlation coefficient for image recognition (the best Pearson correlation coefficient values are highlighted in bold)}
\begin{tabular}{|l|r|r|r|r|} \hline
target&adaptivity&JS-divergence&l2-distance&source-accuracy\\\hline \hline
Balanced&$\textbf{0.93}$&$-0.92$&$-0.62$&$0.80$\\ \hline
Skewed-1&$\textbf{0.86}$&$-0.53$&$-0.52$&$0.78$\\ \hline
Skewed-2&$0.39$&$-\textbf{0.70}$&$-0.66$&$0.07$\\ \hline
Skewed-3&$\textbf{0.25}$&$-0.06$&$-0.18$&$0.73$ \\ \hline
Skewed-4&$\textbf{0.77}$&$-0.41$&$-0.38$&$0.39$\\ \hline
\end{tabular}
\vspace{-5pt}
\end{table}

\begin{table}
\centering
\scriptsize
\caption{\label{tab:comparison-error-rate-image} Comparison of error rate for image recognition}
\begin{tabular}{|l|r|r|r|r|} \hline
&adaptivity&JS-divergence&l2-distance&source-accuracy\\\hline \hline
top-1 error&$\textbf{20\%}$&$40\%$&$40\%$&$60\%$\\ \hline
top-2 error&$\textbf{0\%}$&$0\%$&$0\%$&$40\%$\\ \hline
\end{tabular}
\vspace{-5pt}
\end{table}

\subsubsection{Natural Language Processing}
For this experiment, as mentioned in Sec.~\ref{workloads}, because each task involves merely three datasets, a model search scenario for a target dataset only requires to compare two datasets. Therefore, instead of computing the Pearson coefficient for each search scenario, we choose to compute for each task that contains three search scenarios. For the same reason, we only consider top-$1$ error for this experiment, which is counted over all six search scenarios across the two tasks. 

The results are illustrated in Tab.~\ref{tab:comparison-to-model-adaptivity-text} and Tab.~\ref{tab:comparison-error-rate-image}, which show that while adaptivity has less correlation with the target accuracy compared to JS-divergence, it can more effectively predict the best model for serving with zero error rate for all six search scenarios. The source accuracy performs much worse than other metrics.

\begin{table}
\centering
\scriptsize
\caption{\label{tab:comparison-to-model-adaptivity-text} Comparison of Pearson correlation coefficient for NLP (the best Pearson correlation coefficient values are highlighted in bold)}
\begin{tabular}{|l|r|r|r|} \hline
target&adaptivity&JS-divergence&source-accuracy\\\hline \hline
Task1&$0.61$&$-\textbf{0.71}$&$-0.02$\\ \hline
Task2&$0.76$&$-\textbf{0.87}$&$-0.10$\\ \hline
\end{tabular}
\vspace{-5pt}
\end{table}

\begin{table}
\centering
\scriptsize
\caption{\label{tab:comparison-error-rate-image} Comparison of error rate for NLP}
\begin{tabular}{|l|r|r|r|} \hline
&adaptivity&JS-divergence&source-accuracy\\\hline \hline
top-1 error&$\textbf{0\%}$&$16.7\%$&$33.3\%$\\ \hline
\end{tabular}
\vspace{-5pt}
\end{table}

\subsection{Further Discussion}

\begin{figure}[H]
\centering\subfigure[JS-divergence]{%
   \label{fig:jsd-latency}
   \includegraphics[width=1.58in]{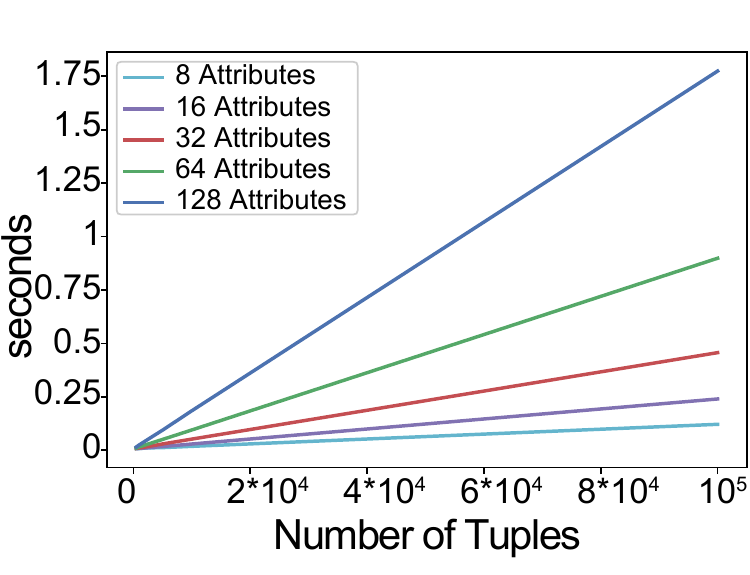}  
}%
\hspace{0pt}
\subfigure[adaptivity]{%
  \label{fig:adaptivity-latency}
  \includegraphics[width=1.58in]{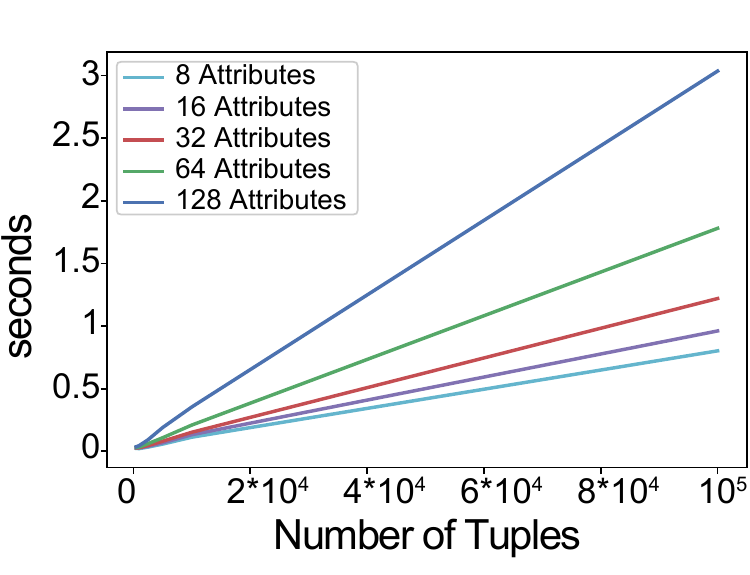}
}%

\vspace{1pt}
\subfigure[source and target both have 2500 rows and 32 cols]{%
  \label{fig:adaptivity-latency}
  \includegraphics[width=1.58in]{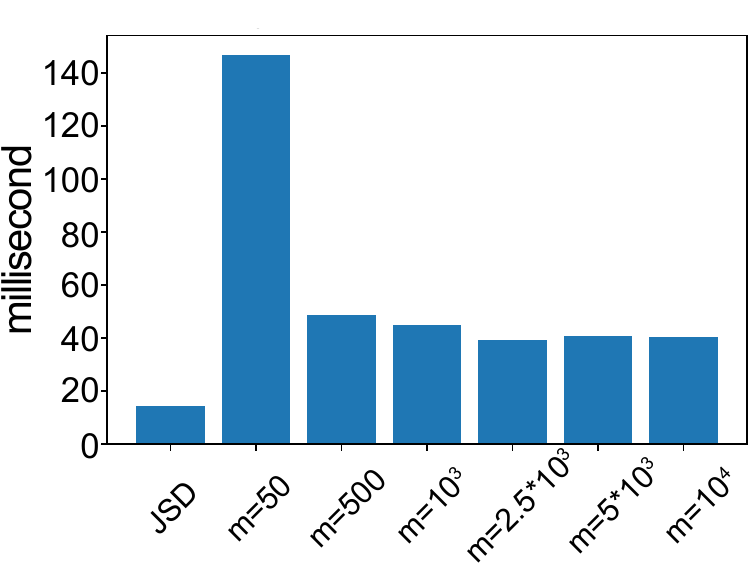}
}%
\hspace{0pt}
\subfigure[source and target both have 10000 rows and 32 cols]{%
  \label{fig:adaptivity-latency}
  \includegraphics[width=1.58in]{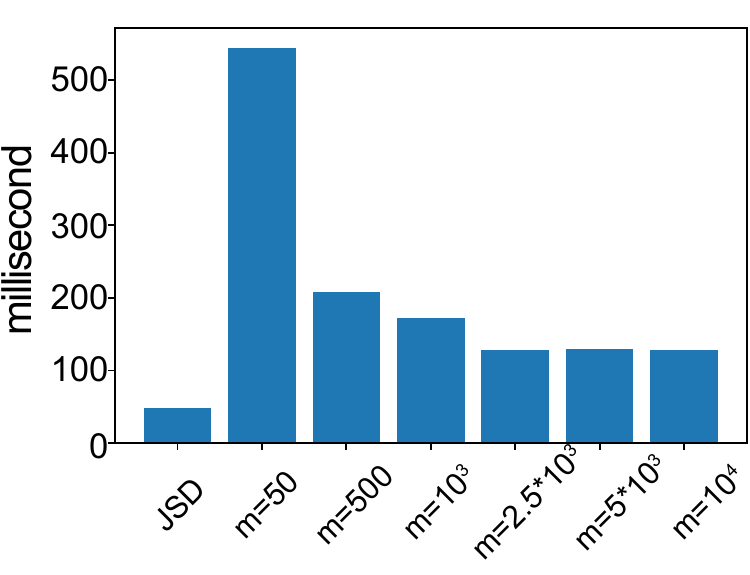}
}%
\eat{
\vspace{1pt}
\subfigure[latency comparison]{%
  \label{fig:adaptivity-latency}
  \includegraphics[width=1.58in]{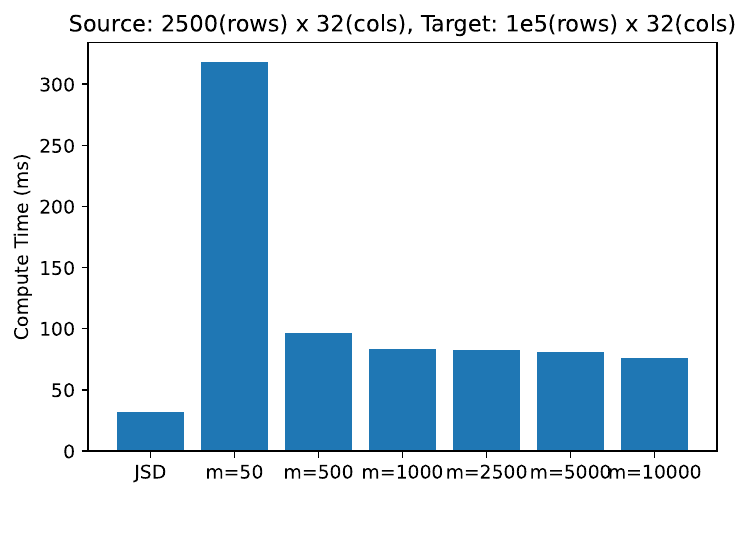}
}%
\hspace{0pt}
\subfigure[latency comparison]{%
  \label{fig:adaptivity-latency}
  \includegraphics[width=1.58in]{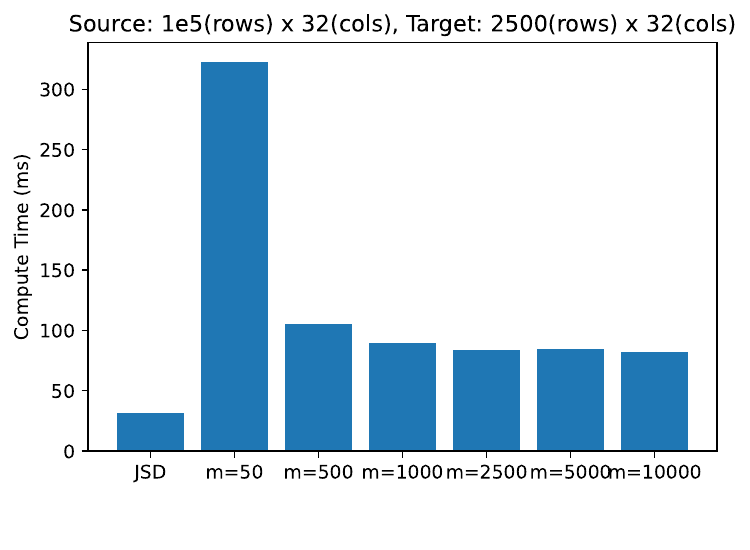}
}%
}

\caption{\label{fig:latency}
Latency comparison of computing JS-divergence and adaptivity, m is the size of each partition
\vspace{-5pt}
}
\end{figure}

One thing to note is that the computation of adaptivity is more complicated than JS-divergence and takes more time. For most of above experiments, we choose the partition size of each dataset between $300$ to $800$. For the NLP experiment, when the source dataset and the target dataset has a significant size discrepancy, we choose the size of the smaller dataset to be the partition size, so that the smaller dataset has only one partition. Then we compare the latency of computing JS-divergence and adaptivity by using different partition sizes and for different sizes of source and target datasets. The results are illustrated in Fig.~\ref{fig:latency}, which illustrates that when partition size is around $500$, the computational overhead for adaptivity is $1.7-3\times$ of JS-divergence.

%% file: relatedworks.tex
\section{Related Works}

In recent, numerous works are proposed to address open data discovery problems, including automatically discover table unionability~\cite{nargesian2018table} and joinability~\cite{zhu2016lsh,zhu2019josie}, and related tables~\cite{zhang2020finding}. Most of these works are trying to identify all similar features using LSH techniques based on Jaccard similarity or variants. While these works are helpful to this study in identifying the overlapping feature space, they are not directly applicable in selecting related models, because the existence of a significant portion of shared features between the source and target datasets doesn't mean the probability distribution of the feature space of the two datasets are similar. 

Zamir and et al~\cite{zamir2018taskonomy} propose a computational approach to find the transferable relationships, abstracted as taskonomy among the different computation vision tasks, e.g. depth estimation, edge detection, point matching and etc., so that some tasks can be trained using other tasks' output and thus require less training data and supervision budget. Their work is focused on the adaptivity from a task's output domain to another task's input domain. In contrast, our work is focused on the adaptivity of tasks' input domains. Also while their work is limited to computer vision, our work is targeting at a more generalized scenario. Kornblith and et al~\cite{kornblith2019better} study the relationship between the ImageNet model architectures such as ResNet, MobileNet, Inception-ResNet, etc., and the transfer learning performance. They observe that transfer learning based on finetune techniques (i.e., just tune the last two layers) even using a "better" model architecture may not achieve "better" accuracy. This means in regard of the effectiveness of transfer learning, the divergence of features between the source and target domains is more important than the difference of model architectures. Their observation indicates that our work may be applicable to find related models for transfer learning.